\documentclass{article}

\pdfoutput=1
\usepackage{Others/PRIMEarxiv}

\usepackage[utf8]{inputenc} 
\usepackage[T1]{fontenc}    
\usepackage{url}            
\usepackage{booktabs}       
\usepackage{amsfonts}       
\usepackage{amsmath}
\usepackage{amssymb}
\usepackage{mathtools}

\DeclareMathOperator*{\argmin}{arg\,min}
\usepackage{algorithm}
\usepackage{algpseudocode}

\usepackage{lscape}
\usepackage{nicefrac}       
\usepackage{microtype}      
\usepackage{lipsum}
\usepackage{fancyhdr}       
\usepackage{graphicx}       
\usepackage{xcolor}
\usepackage{setspace}
\usepackage{chngcntr}

\usepackage[
    natbib=true,
    style=numeric-comp,
    sorting=none
]{biblatex}
\usepackage[pdftex]{hyperref}

\title{\vspace{5mm}\fontsize{18pt}{20pt}\selectfont\textbf{Physics reliable frugal uncertainty analysis for full waveform inversion}} 

\author{
\large
\textsc{M. Izzatullah, M. Ravasi, T. Alkhalifah}\\[2mm]
\large DeepWave Consortium;\\ King Abdullah University of Science and Technology, Thuwal, Kingdom of Saudi Arabia. 
}
\date{}

\begin{document}
\maketitle
\doublespacing

\begin{abstract}
Full waveform inversion (FWI) enables us to obtain high-resolution velocity models of the subsurface. However, estimating the associated uncertainties in the process is not trivial. Commonly, uncertainty estimation is performed within the Bayesian framework through sampling algorithms to estimate the posterior distribution and identify the associated uncertainty. Nevertheless, such an approach has to deal with complex posterior structures (e.g., multimodality), high-dimensional model parameters, and large-scale datasets, which lead to high computational demands and time-consuming procedures. As a result, uncertainty analysis is rarely performed, especially at the industrial scale, and thus, it drives practitioners away from utilizing it for decision-making. This work proposes a frugal approach to estimate uncertainty in FWI through the Stein Variational Gradient Descent (SVGD) algorithm by utilizing a relatively small number of velocity model particles. We warm-start the SVGD algorithm by perturbing the optimized velocity model obtained from a deterministic FWI procedure with random field-based perturbations. Such perturbations cover the scattering (i.e., high wavenumber) and the transmission (i.e., low wavenumber) components of FWI and, thus, represent the uncertainty of the FWI holistically. We demonstrate the proposed approach on the Marmousi model; we have learned that by utilizing a relatively small number of particles, the uncertainty map presents qualitatively reliable information that honours the physics of wave propagation at a reasonable cost, allowing for the potential for industrial-scale applications. Nevertheless, given that uncertainties are underestimated, we must be careful when incorporating them into downstream tasks of seismic-driven geological and reservoir modelling.
\end{abstract}
\vspace{0.5cm}
\section{Introduction}
Full waveform inversion (FWI) is a high-resolution seismic imaging technique commonly used to estimate subsurface velocity models from full wavefield seismic data~\citep[see, e.g.,][]{tarantola_inversion_1984,tarantola_strategy_1986,gauthier_twodimensional_1986,virieux_overview_2009,metivier_full_2017}. Since FWI is a highly nonlinear, ill-posed inverse problem, estimating the associated uncertainties is vital for any subsequent decision-making process. However, such a task is computationally demanding and time-consuming due to the complex posterior structure (e.g., multimodality), high-dimensional model parameters, and large-scale datasets~\citep[see, e.g.,][]{tarantola_generalized_1982,mosegaard_monte_1995}. Thus, it drives the practitioners away from utilizing it for decision-making, especially in the industrial-scale FWI.\\

Over the past decade, the uncertainties estimation (resolution analysis included), especially in FWI, is often reduced, for example, to the Laplace approximation approach~\citep[e.g.,][]{fichtner_resolution_2011, fichtner_resolution_2015, fang_uncertainty_2018,liu_square-root_2019,izzatullah_bayesian_2019, izzatullah_visualizing_2019, izzatullah_bayes_2021, izzatullah_laplace_2022} to alleviate such limitations, especially on the computational costs. Although the Laplace approximation is the simplest family of posterior approximations for high-dimensional problems (e.g., by forming a Gaussian approximation to the exact posterior in the locality of the maximum a posteriori (MAP) solution), in the case of FWI, it requires numerous computationally intensive forward and adjoint operator evaluation per point of model discretization to approximate its covariance matrix. In this case, the Laplace approximation still faces computational limitations in estimating uncertainties in FWI at the industrial scale.\\

Another approach for uncertainty estimation is the Markov chain Monte Carlo (MCMC) method. MCMC is considered, in general, the gold standard in Bayesian inference to describe a target posterior distribution~\citep{murphy_probabilistic_2023}. In the geophysics community, robust and efficient gradient-based Markov Chain Monte Carlo (MCMC) methods are being explored as alternatives to estimate uncertainties in FWI. For example, the Langevin Monte Carlo (LMC)~\citep{izzatullah_bayesian_2020, siahkoohi_uncertainty_2020, izzatullah_bayesian_2021, izzatullah_langevin_2021, izzatullah_approximate_2021} and Hamiltonian Monte Carlo (HMC)~\citep[e.g.,][]{sen_transdimensional_2017, fichtner_hamiltonian_2019, fichtner_hamiltonian_2019-1, gebraad_bayesian_2020, fichtner_autotuning_2021, zunino_hamiltonian_2022} are used in Bayesian FWI to sample from high-dimensional posterior distributions. Despite their robustness and efficiency, these MCMC methods require a large number of sampling steps (e.g., on the order of $\sim10^{4}$ and above) and a long burn-in period to perform an accurate uncertainties estimation due to their sequential nature (i.e., Markov chain property), which hinders their applicability and scalability for industrial-scale uncertainty estimation, for FWI, due to the costs associated with the forward and adjoint operators evaluation at each iteration.\\

As an alternative to the conventional approaches mentioned above (i.e., Laplace approximation and MCMC methods) in uncertainty estimation, especially for FWI, the variational inference (VI) approach would be an optimal choice~\citep{blei_variational_2017, zhang_advances_2018}. VI provides a balance between the Laplace approximation and MCMC methods because it provides more flexibility in approximating the posterior distributions at a reduced computational cost, especially in high-dimensional large-scale inverse problems such as FWI~\citep[e.g.,][]{nawaz_variational_2020, urozayev_reduced-order_2022}. The particles update approach, such as Stein Variational Gradient Descent (SVGD), is one of the VI algorithms that has shown success in performing large-scale Bayesian inference in FWI~\cite[e.g.,][]{zhang_variational_2020, zhang_introduction_2021, zhang_3d_2023}. Nevertheless, SVGD still suffers from high computational costs as it requires a large number of particles to sample the whole posterior. Therefore, instead of describing the whole posterior distribution, SVGD could be utilized for estimating the uncertainty locally using a much smaller number of particles. Arguably, this approach compromises, statistically, the strictness of Bayesian inference as it provides biased analysis due to the limited number of particles. However, such an approach in FWI at the industrial scale is promising as it saves computational costs while practically assisting decision-making with qualitatively relevant information.\\

An essential requirement for an uncertainty evaluation in FWI that properly represents the physics is that the prior distribution in VI, or the perturbations of the initial SVGD particles, should represent the transmission and scattering components of FWI. Claerbout in 1985~\citep{claerbout_imaging_1985} highlighted the sensitivity of the data to various scales of the velocity model in a diagram that divided the sensitivity into a background (i.e., low wavenumber) scale that controls wave propagation and a scattering (i.e., high wavenumber) scale that induces reflections, with a gap in between. Peter Mora in 1989~\cite{mora_inversion_1989} described these two components in the context of FWI in terms of tomography and migration. An in-depth analysis of the relationship between these scales and the FWI machinery can be found in~\citep{alkhalifah_full-model_2016, alkhalifah_multiscattering_2016}. To quantify uncertainty representative of these model scales (i.e., wavenumbers) in VI, the prior and initial SVGD particles should reflect these scales. So, in this study, we investigate various options to do so for the FWI inverse problem.\\

Thus, we propose a frugal approach to estimate uncertainty in FWI through the SVGD algorithm by utilizing a relatively small number of velocity model particles. By doing so, we can provide uncertainty maps (i.e., relative standard deviation maps) at a much lower cost, allowing for potential industrial-scale applications. We also introduce an implementation of SVGD in which its initial particles are sampled by perturbing the inverted velocity model from the deterministic FWI optimization using random field perturbations as a warm start. Unlike previously used uncorrelated random perturbations, such as Gaussian noise and uniformly distributed perturbations, the random field perturbations allow us to inject both the transmission and scattering components of FWI into this uncertainty estimation. As a result, the obtained uncertainty map presents qualitatively reliable information that honours the physics of wave propagation and reflects the uncertainty we would expect from FWI.\\

The rest of the manuscript is organized as follows: in Section 2, for completeness, we describe the theoretical framework based on the Bayesian framework for FWI, followed by providing an overview of the VI, the SVGD algorithm, and Gaussian Random Field (GRF). Section 3 highlights the practicality of our proposed frugal approach with numerical examples based on the Marmousi model. Finally, we discuss the proposed approach's potential and limitations in Section 4 and finally share our conclusions.\\
\section{Theoretical Framework}
This work aims to establish a frugal uncertainty estimation framework for FWI, which allows for potential industrial-scale applications. Here, we leverage the SVGD~\citep{liu_stein_2016}, an algorithm within the VI approach, by utilizing a relatively small number of velocity model particles. As mentioned in the introduction, the approach we propose here compromises the strictness of the Bayesian formulation by deriving relevant statistical information about the uncertainty in FWI through a relatively small set of samples, considering the dimension of the problem. As a result, the statistics obtained from this approach are biased. Also, note that in this work, we focus on estimating the uncertainty map (i.e., relative standard deviation), which informs us of the credibility of subsurface regions inverted by FWI as it relates to the physics of wave propagation, considering the experiment at hand, instead of describing the full posterior distribution analysis as commonly performed in Bayesian inference. Nevertheless, this section briefly overviews the Bayesian and deterministic formulations for FWI and the SVGD algorithm. We also share the most effective perturbation models in the context of SVGD on FWI.\\

\subsection{Bayesian FWI formulation} \label{sec:theory-bayes}
In FWI, we seek to estimate the subsurface model parameters (e.g., seismic velocity or slowness) $\mathbf{m} \in \mathbb{R}^{m}$ from the observed seismic data $\mathbf{d} \in \mathbb{R}^{d}$, where $m$ and $d$ are the dimensions of the model parameters and the observed seismic data, respectively. The uncertainty estimation in FWI is commonly formulated based on the Bayesian formulation, which can be formulated through Bayes' Theorem~\cite[ch.~3.1.2]{murphy_probabilistic_2023}, where the posterior ($\mathbf{p}(\mathbf{m}|\mathbf{d})$) can be defined up to its normalizing constant as
\begin{equation}\label{eq:unnormalized-bayes}
    \pi(\mathbf{m}) \triangleq \mathbf{p}(\mathbf{m}|\mathbf{d}) = \frac{1}{Z}\mathbf{p}(\mathbf{d}|\mathbf{m})\mathbf{p}(\mathbf{m}).
\end{equation}
where $Z$ is the normalization constant that is independent of $\mathbf{m}$. The Bayesian formulation stated above introduces the notion of prior probability density ($\mathbf{p}(\mathbf{m})$) and the likelihood function  ($\mathbf{p}(\mathbf{d}|\mathbf{m})$). Here, the prior probability density encodes the confidence of the prior information of the unknown seismic velocity model $\mathbf{m}$. In contrast, the likelihood function describes the conditional probability density of the seismic velocity model to model the seismic data. Based on Bayes' theorem, we obtain the posterior probability density of the seismic velocity model given the observed data, $\mathbf{p}(\mathbf{m}|\mathbf{d})$, by combining the prior probability density and the likelihood function. 

\subsection{Deterministic FWI formulation}
For readers familiar with the deterministic formulation of FWI, the Bayesian formulation above, overall, can be regarded as a more general framework for performing FWI. Bayesian FWI relates to deterministic FWI by solving the optimization problem where the maximum-a-posteriori (MAP) solution is sought. Equation~\eqref{eq:unnormalized-bayes} can be reformulated in a deterministic setting as a minimization of the negative log-posterior distribution as follows:
\begin{equation}\label{eq:log-posterior-minimization}
\begin{split}
    \mathbf{m}^{\ast} &= \argmin_{\mathbf{m}} -\log\pi(\mathbf{m})\\
    &= \argmin_{\mathbf{m}} -\log\mathbf{p}(\mathbf{d}|\mathbf{m}) - \log\mathbf{p}(\mathbf{m}) + \log Z\\
    &= \argmin_{\mathbf{m}} -\underbrace{\log\mathbf{p}(\mathbf{d}|\mathbf{m})}_{\text{data-misfit term}} - \underbrace{\log\mathbf{p}(\mathbf{m})}_{\text{regularization term}},
\end{split}
\end{equation}
where the constant term $\log Z$ in the above equation can be ignored without affecting the inference/optimization procedure. \\

The common choice for the log-likelihood (data-misfit) term above for FWI is the l2-norm~\citep{tarantola_generalized_1982, tarantola_strategy_1986,mora_nonlinear_1987, mora_inversion_1989}, which often provides the highest resolution model information and is often applied at the end of any FWI application, especially with any sophisticated data-misfit~\citep[see, e.g.,][]{bozdag_misfit_2011, engquist_optimal_2016, guasch_adaptive_2019, li_extended_2021, pladys_cycle-skipping_2021}. The l2-norm misfit is associated with the Gaussian errors in the data and can be described as
\begin{equation}\label{eq:log-likelihood}
    \log\mathbf{p}(\mathbf{d}|\mathbf{m}) = -\frac{1}{2\sigma^{2}}\|\mathbf{F}(\mathbf{m}) - \mathbf{d}\|^{2},
\end{equation}
where $\sigma$ represents the data standard deviation and $\mathbf{F}(\mathbf{m})$ is a nonlinear forward operator that maps the velocity model $\mathbf{m} \in \mathbb{R}^{m}$ onto the observed seismic data $\mathbf{d} \in \mathbb{R}^{d}$. The forward operator $\mathbf{F}(\mathbf{m})$ is highly nonlinear with respect to the velocity model $\mathbf{m}$. This log-likelihood term measures the information contributed from observed data in reconstructing the seismic velocity model. It also encompasses all the information regarding the physics of wave propagation in FWI, which relates to both high and low wavenumbers information updates~\citep[see, e.g.,][]{mora_nonlinear_1987, mora_elastic_1988, mora_inversion_1989}.\\

Meanwhile, the log-prior, a.k.a regularization term, provides additional prior information, apart from the data, about the seismic velocity model, e.g., roughness and smoothness of subsurface structures, into FWI by the practitioners. Such additional information affects the shape of the posterior distribution $\pi(\mathbf{m})$, thus, influencing the statistics for uncertainty analysis and the performance of sampling/optimization methods, respectively. Imposing proper prior information (regularization) on the posterior distribution (log-posterior) may also improve the credibility of FWI results.\\
%
\subsection{Variational inference (VI) and Stein Variational Gradient Descent (SVGD)}\label{sec:theory-vi}
In comparison to the deterministic FWI or other conventional inference approaches, such as the Laplace approximation and MCMC, the VI approach approximates the posterior distribution through a surrogate distribution $q(\mathbf{m})$ that is easy to sample (e.g., Gaussian distribution). VI performs the inference procedure through a functional minimization problem of the KL divergence between the surrogate $q(\mathbf{m})$ and a target posterior distributions $\pi(\mathbf{m})$~\cite{blei_variational_2017, zhang_advances_2018}. The KL divergence is given by
\begin{equation} \label{eq:kl-div}
    \mathrm{KL}(q(\mathbf{m})||\pi(\mathbf{m})) = \mathrm{E}_{\mathbf{m} \sim q}[-\log\pi(\mathbf{m}) + \log q(\mathbf{m})],
\end{equation}
where in the right hand side of ~\eqref{eq:kl-div} above, we evaluate the expectation of $\mathbf{m}$ sampled from the surrogate distribution $q(\mathbf{m})$. By minimizing the equation \eqref{eq:kl-div}, VI approximates a target posterior distribution $\pi(\mathbf{m})$ by solving
\begin{equation} \label{eq:min-kl-vi}
\begin{split}
    q^{\ast} &= \argmin_{q} \mathrm{E}_{\mathbf{m} \sim q}[-\log\pi(\mathbf{m}) + \log q(\mathbf{m})]\\
    &= \argmin_{q} \mathrm{KL}(q(\mathbf{m})||\pi(\mathbf{m})).
\end{split}
\end{equation}
Here, different choices of $q(\mathbf{m})$ give rise to different VI methods. One particular VI method formulated based on the particles evolution approach to solve equation~\eqref{eq:min-kl-vi} above is called Stein variational gradient descent (SVGD)~\cite{liu_stein_2016}.\\

SVGD starts from an arbitrary initial density $q_{0}$ and the associated particles $\mathbf{m}_{i, 0} \sim q_{0}$, where $\mathbf{m}_{i, 0}$ is a $m$-dimensional vector represents a seismic velocity model. These particles are updated through a smooth transform $\mathbf{T}(\mathbf{m})=\mathbf{m} + \eta\phi(\mathbf{m})$, where $\phi(\mathbf{m})$ is a smooth function that describes the evolution direction, similar to a gradient direction in the FWI setting, and the scalar $\eta$ represents the step size. We assume the transformation $\mathbf{T}$ is invertible and define $q_{[\mathbf{T}]}(\mathbf{m})$ as the transformed density by the transformation $\mathbf{T}$.\\

Based on Stein's method, the gradient of the KL divergence between $q_{[\mathbf{T}]}(\mathbf{m})$ and a target posterior $\pi(\mathbf{m})$ with respect to the step size $\eta$ can be described through the \emph{Stein identity} as:
\begin{equation} \label{eq:grad-kl}
    \nabla_{\varepsilon}\mathrm{KL}(q_{[\mathbf{T}]}(\mathbf{m})||\pi(\mathbf{m}))|_{\varepsilon=0} = -\mathrm{E}_{\mathbf{m} \sim q}[\text{trace}(\mathcal{A}_{\pi}\phi(\mathbf{m}))],
\end{equation}
where $\mathcal{A}_{\pi}\phi(\mathbf{m}) = \nabla_{\mathbf{m}}\log\pi(\mathbf{m})\phi(\mathbf{m}) + \nabla_{\mathbf{m}}\phi(\mathbf{m})$ is the \emph{Stein operator} \cite{stein_bound_1972, gorham_measuring_2015, gorham_measuring_2017, gorham_measuring_2019, izzatullah_bayesian_2020}. Here, equation \eqref{eq:grad-kl} provides a form of functional gradient descent direction of the KL divergence. Therefore, the KL divergence can be minimized in an iterative manner by a small magnitude of $\eta$. To minimize the KL divergence as fast as possible, the function $\phi(\mathbf{m})$ should be chosen from a set of functions such that the decreasing rate of $-\frac{d}{dt}(q_{[\mathbf{T}]}(\mathbf{m})||\pi(\mathbf{m}))$ is maximized. Typically, this set of functions is chosen to be positive definite with continuously differentiable kernel functions from the \emph{reproducing kernel Hilbert space (RKHS)}~\cite{berlinet_reproducing_2011, paulsen_introduction_2016}. By the reproducing property of RKHS and Stein identity in equation \eqref{eq:grad-kl}, we can show that the optimal $\phi^{\ast}$ is
\begin{equation}\label{eq:optimal-phi}
    \phi^{\ast}(\cdot) = \mathrm{E}_{\mathbf{m} \sim q}[\nabla_{\mathbf{m}}\log\pi(\mathbf{m})k(\mathbf{m}, \cdot) + \nabla_{\mathbf{m}}k(\mathbf{m}, \cdot)],
\end{equation}
where $k(\mathbf{m}, \cdot)$ is a RKHS function. Through equation \ref{eq:optimal-phi}, the SVGD of \cite{liu_stein_2016} can be derived as
\begin{equation}\label{eq:svgd}
\begin{split}
    \mathbf{m}_{i, t+1} &= \mathbf{m}_{i, t} + \eta_{t}\mathrm{E}_{\mathbf{m}_{j, t} \sim q_{t}}[\nabla_{\mathbf{m}_{j, t}}\log\pi(\mathbf{m}_{j, t})k(\mathbf{m}_{j, t}, \mathbf{m}_{i, t}) + \nabla_{\mathbf{m}_{j, t}}k(\mathbf{m}_{j, t}, \mathbf{m}_{i, t})]\\
    &= \mathbf{m}_{i, t} + \frac{\eta_{t}}{N}\sum^{N}_{j=1}[\nabla_{\mathbf{m}_{j, t}}\log\pi(\mathbf{m}_{j, t})k(\mathbf{m}_{j, t}, \mathbf{m}_{i, t}) + \nabla_{\mathbf{m}_{j, t}}k(\mathbf{m}_{j, t}, \mathbf{m}_{i, t})]\\
    &= \mathbf{m}_{i, t} + \eta_{t}\phi^{\ast}_{t}(\mathbf{m}_{i, t}),
\end{split}
\end{equation}
for $N$ particles $\{\mathbf{m}_{i,t}\}_{i=1}^{N} \in \mathrm{R}^{m}$. We summarize the SVGD procedure in Algorithm~\ref{alg:svgd}. SVGD forms a natural counterpart of gradient descent for optimization by iteratively updating those particles from their initial density $q_{0}(\mathbf{m})$ to match the target posterior $\pi(\mathbf{m})$ by minimizing the KL divergence. Also, SVGD has found great success in Bayesian FWI~\cite{zhang_variational_2020, zhang_introduction_2021, zhang_3d_2023}.\\

The advantage of SVGD is that its initial particles can be sampled from an arbitrary distribution. We thus leverage this concept in this work by perturbing the inverted velocity model from the deterministic FWI optimization by random field-based perturbations and warm start SVGD by considering them as the initial particles. Note that, the perturbations along this inverted velocity model lie within the respective posterior distribution, i.e., in the vicinity of the optimal local minimum among the multimodality distribution of FWI. In our case, these initial particles belong to one of the distribution modes due to the multimodality nature of FWI. This approach allows us only to quantify the uncertainties locally within the mode's proximity. However, it does not address uncertainties related to whether the model or part of it has converged to a local minimum solution. Thus, we assume that the inverted model in which we perturb from locally is the optimal solution, at least in most of its dimensions. Nevertheless, the relevant information obtained from those particles within this region will practically and qualitatively assist the practitioners in decision-making tasks.
%
\begin{algorithm}
\caption{Stein Variational Gradient Descent (SVGD)}\label{alg:svgd}
\textbf{Input:} A target posterior distribution $\pi(\mathbf{m})$ and a set of initial particles $\{\mathbf{m}_{i,t}\}_{i=1}^{N}$.\\
\textbf{Output:} A set of particles $\{\mathbf{m}_{i,t}\}_{i=1}^{N}$ that approximates the target posterior distribution.
\begin{algorithmic}
\For{iteration $t$} 

\State {$\mathbf{m}_{i,t+1} \longleftarrow \mathbf{m}_{i,t} - \eta_{t}\phi^{\ast}_{t}(\mathbf{m}_{i, t})$ where $\phi^{\ast}(\mathbf{m}) = \frac{1}{N}\sum^{N}_{j=1}[\nabla_{\mathbf{m}_{j, t}}\log\pi(\mathbf{m}_{j, t})k(\mathbf{m}_{j, t}, \mathbf{m}) + \nabla_{\mathbf{m}_{j, t}}k(\mathbf{m}_{j, t}, \mathbf{m})]$}

\State{where $\eta_{t}$ is the step size at the $t$-th iteration.}

\EndFor
\end{algorithmic}
\end{algorithm}

\subsection{SVGD perturbation models for FWI uncertainty estimation}
Previously, we briefly discussed the FWI formulation. According to~\citep{mora_inversion_1989}, FWI is equivalent to doing migration and reflection tomography simultaneously. By such definition, FWI consists of two major components, the scattering (i.e., high wavenumber) and transmission (i.e., low wavenumber) components, where both are simultaneously being updated in the FWI procedure. Furthermore, these two components are associated with the log-likelihood (data-misfit) term, thus, becoming a source of uncertainties in FWI that needs to be estimated. In order to capture the associated uncertainties, the perturbation (prior) models used in the uncertainty estimation procedure play a major role.\\

To our knowledge, most of the literature on uncertainty estimation for FWI only demonstrated the uncertainty with respect to the scattering component through the additive uncorrelated noise perturbation model, such as Gaussian noise and uniformly distributed perturbations. Such a perturbation model introduces random scattering to the process, thus mainly representing the uncertainty in the scattering (i.e., high wavenumber) component of the FWI, which is closer to what we experience in seismic imaging. In other words, this perturbation has almost zero variance with respect to the long wavelength components of the model~\citep[see, e.g.,][]{gebraad_bayesian_2020, izzatullah_bayesian_2021, zhang_introduction_2021, zhang_3d_2023, izzatullah_frugal_2023}. Meanwhile, the transmission (i.e., low wavenumber) component, which controls wave propagation, has not been addressed. Therefore, we provide a discussion on addressing the uncertainty related to this component in the discussion section.\\

In this work, we demonstrate the uncertainty estimation based on the contribution of both scattering and transmission components of FWI. To estimate the uncertainty derived from such contributions, we rely on generating Gaussian Random Field (GRF) perturbation model. The motivation behind the usage of the random fields-based perturbation model, especially GRF, is that it has a natural physical interpretation, providing randomness in amplitude and scale of the model, which covers the range of model wavenumbers producing data sensitivities corresponding to the transmission and scattering components of FWI.\\


Here, we consider realizations of the GRF perturbations as GRF with Mat\'ern covariance kernel function $\mathbf{C}_{\nu}$~\citep[e.g.,][]{bogachev_gaussian_1998, lord_introduction_2014, ghosal_fundamentals_2017} such that
\begin{equation}\label{eq:grf}
    \mathbf{C}_{\nu}(d) = \sigma^{2} \frac{2^{1-\nu}}{\mathbf{\Gamma}(\nu)}\big(\sqrt{2\nu}\frac{d}{\alpha}\big)^{\nu}\mathbf{K}_{\nu}\big(\sqrt{2\nu}\frac{d}{\alpha}\big),
\end{equation}
where $\sigma$ is the variance of the Gaussian process, $\mathbf{\Gamma}$ is the gamma function~\citep{artin_gamma_2015}, $\mathbf{K}_{\nu}$ is the modified Bessel function of the second kind~\citep{arfken_mathematical_2012}, $\alpha$ is a positive parameter, and $\nu$ is the smoothness parameter of the random field. The distance $\frac{d}{\alpha}$ between two points $\mathbf{m} = (m_{1}, \dots, m_{m})$ and $\mathbf{m^{'}} = (m^{'}_{1}, \dots, m^{'}_{m})$, is defined such that
\begin{equation}\label{eq:grf-distance}
    \frac{d}{\alpha} = \sqrt{\sum_{i=1}^{m}\Big(\frac{m_{i} - m^{'}_{i}}{\lambda_{i}}\Big)^{2}}.
\end{equation}
Here, $\pmb{\lambda}=(\lambda_{1}, \dots, \lambda_{m})$ is the vector coefficient, which defines the correlation length along the two points. This GRF perturbation model provides us with perturbations, as illustrated in Figure~\ref{fig:rf-init}, that consists of a holistic effect on the physics of wave propagation and later contribute to a physics-reliable uncertainty estimation of FWI.\\
\section{Numerical Examples}
This section demonstrates the practicality of our proposed approach, by frugally performing uncertainty estimation in FWI. We present numerical examples based on the Marmousi model. These Marmousi model examples allow us to qualitatively investigate and analyze the uncertainty of FWI through SVGD with a small set of particles. The observed data are modelled using 30 shots with a source interval of 300 m and 300 receivers with an interval of 30 m. The source is given by a Ricker wavelet with a 6 Hz peak frequency. The maximum recording time is 4 s. Next, we add coloured noise to the observed data within the same frequency spectrum. The resulting noisy observed data have a signal-to-noise ratio (SNR) of 11.64 dB. We use only the log-likelihood term in equation~\eqref{eq:log-likelihood} for estimating uncertainty in these examples. We want to investigate the uncertainty derived from the information in the observed data without any additional information, i.e., without the contributions from the log-prior (regularization) term.\\ 

\begin{figure*}[!htb]
  \centering
  \includegraphics[width=\textwidth]{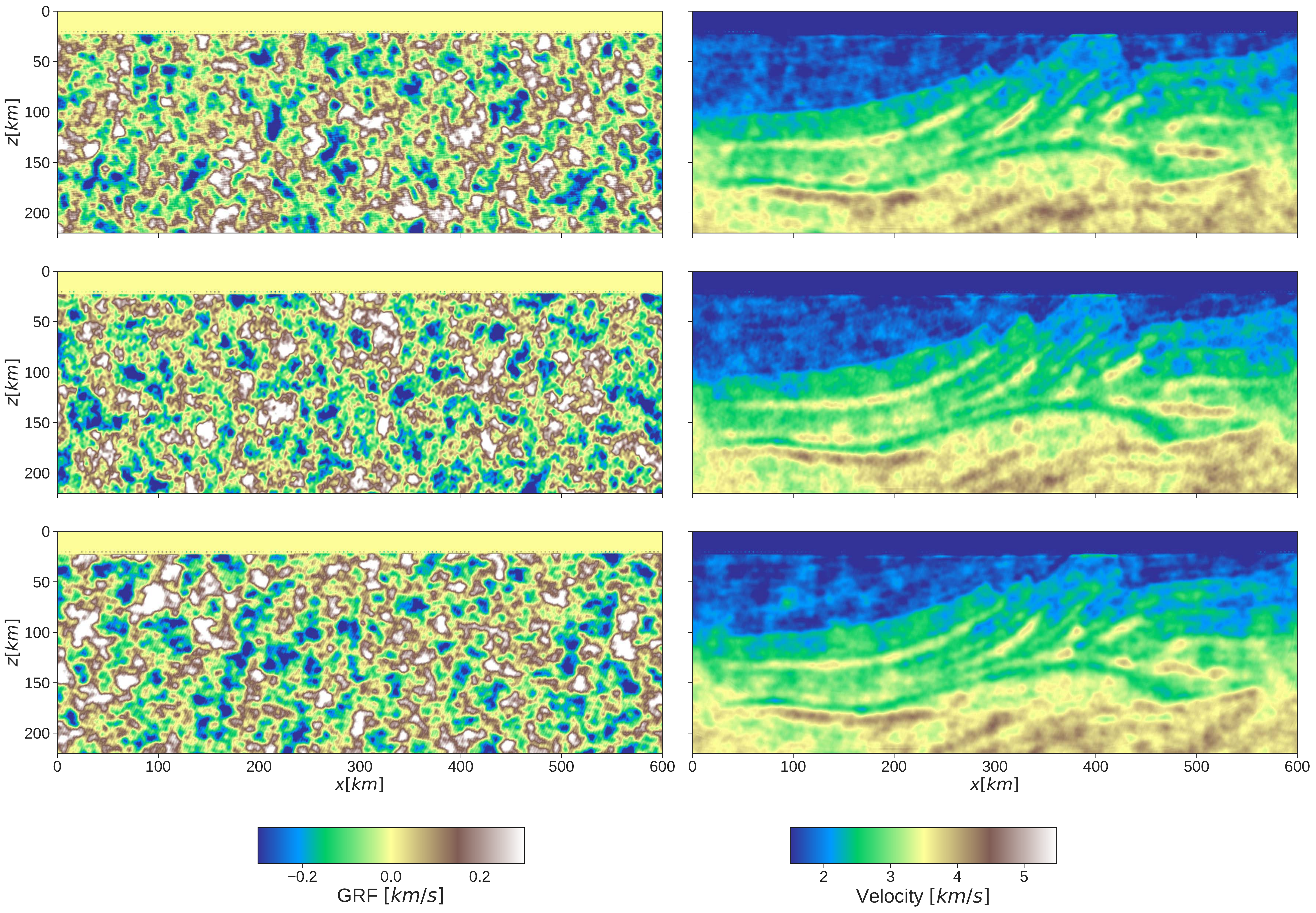}
  \caption{First three realizations of GRF perturbations (left column) and initial SVGD particles with GRF perturbations (right column).}
  \label{fig:rf-init}
\end{figure*}

Here, we perform two simulations with 50 and 100 SVGD particles to evaluate the statistics dependency on the number of particles. For the SVGD algorithm, we use the RBF kernel with bandwidth chosen by the standard median trick, that is, we use $k_{t}(\mathbf{m}, \mathbf{m}^{'}) = \exp({-\|\mathbf{m} - \mathbf{m}^{'}\|^{2} / \sigma_{t}^{2}})$, where the bandwidth $\sigma_{t}$ is set by $\sigma_{t} = \textbf{Median}\{\|\mathbf{m}_{i, t} - \mathbf{m}_{j,t}\|:i \neq j\}$ based on the particles at the $t$-th iteration. We generated the GRF perturbations within a range of $\pm 300$ km/s with parameter $\pmb{\lambda}=0.1$, and the smoothness coefficient is set to be $\nu = 1.25$. We added them to the inverted model from the deterministic FWI procedure with the abovementioned setting to produce the initial SVGD particles, as illustrated in Figure~\ref{fig:rf-init}. Such choices of the initial particles would save computational cost and assist in estimating the uncertainties within the basin of attraction (i.e., one of the distribution modes). We perform 100 iterations of SVGD and collect the updated particles as updated samples for uncertainty analysis at the end of the simulations. These numerical examples have been implemented using PyTorch~\citep{paszke_automatic_2017} and Deepwave~\citep{richardson_alan_2022} on a NVIDIA A100 GPU.\\

\begin{figure*}[!htb]
  \centering
  \includegraphics[width=\textwidth]{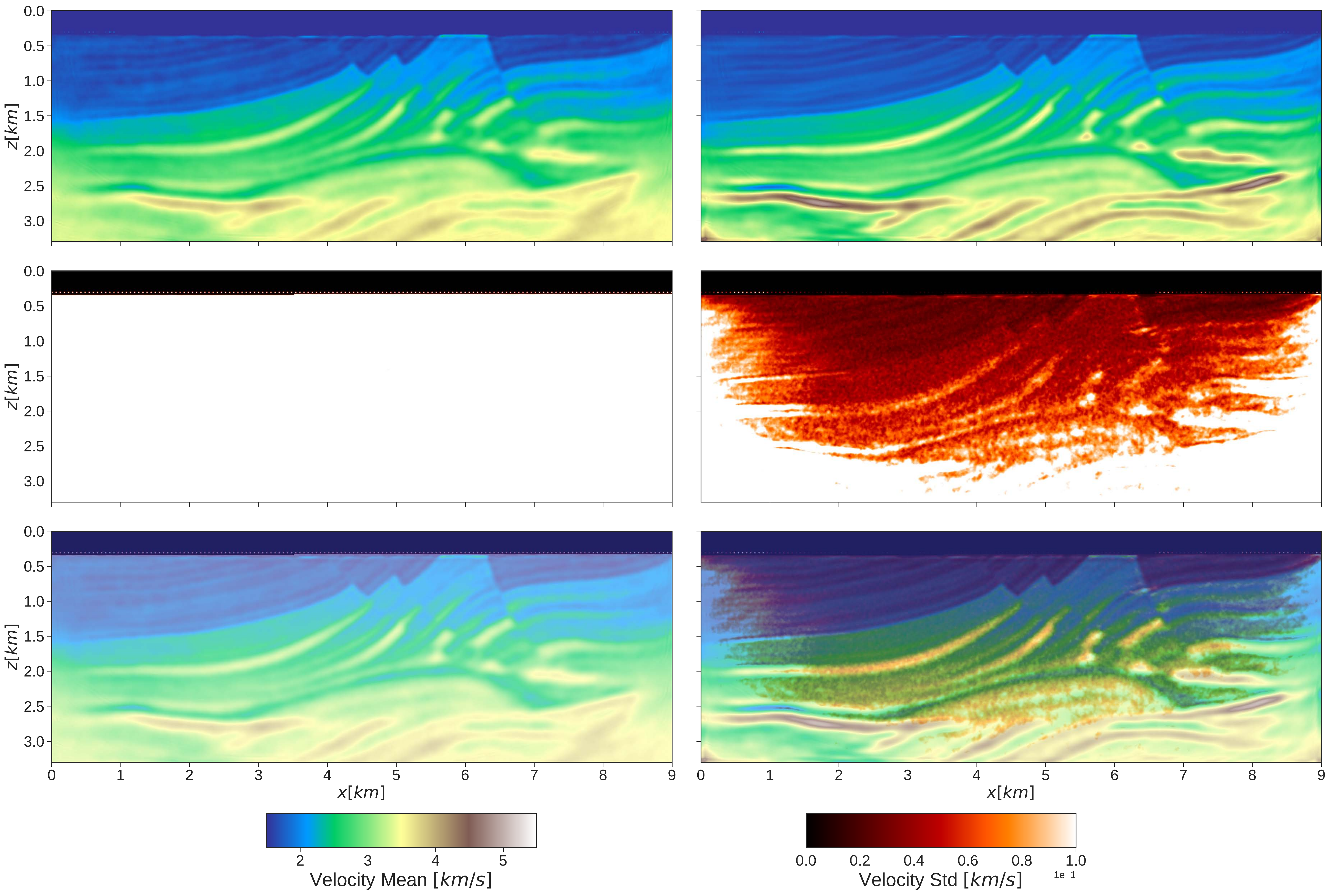}
  \caption{Marmousi model frugal statistics with 50 particles. \textbf{Top row:} Initial and updated particles means. \textbf{Middle row:} Initial and updated particles standard deviations. \textbf{Bottom row:} Overlay of initial and updated particles means on their standard deviations, respectively.}
  \label{fig:rf-stats-50-particle}
\end{figure*}

\begin{figure*}[!htb]
  \centering
  \includegraphics[width=\textwidth]{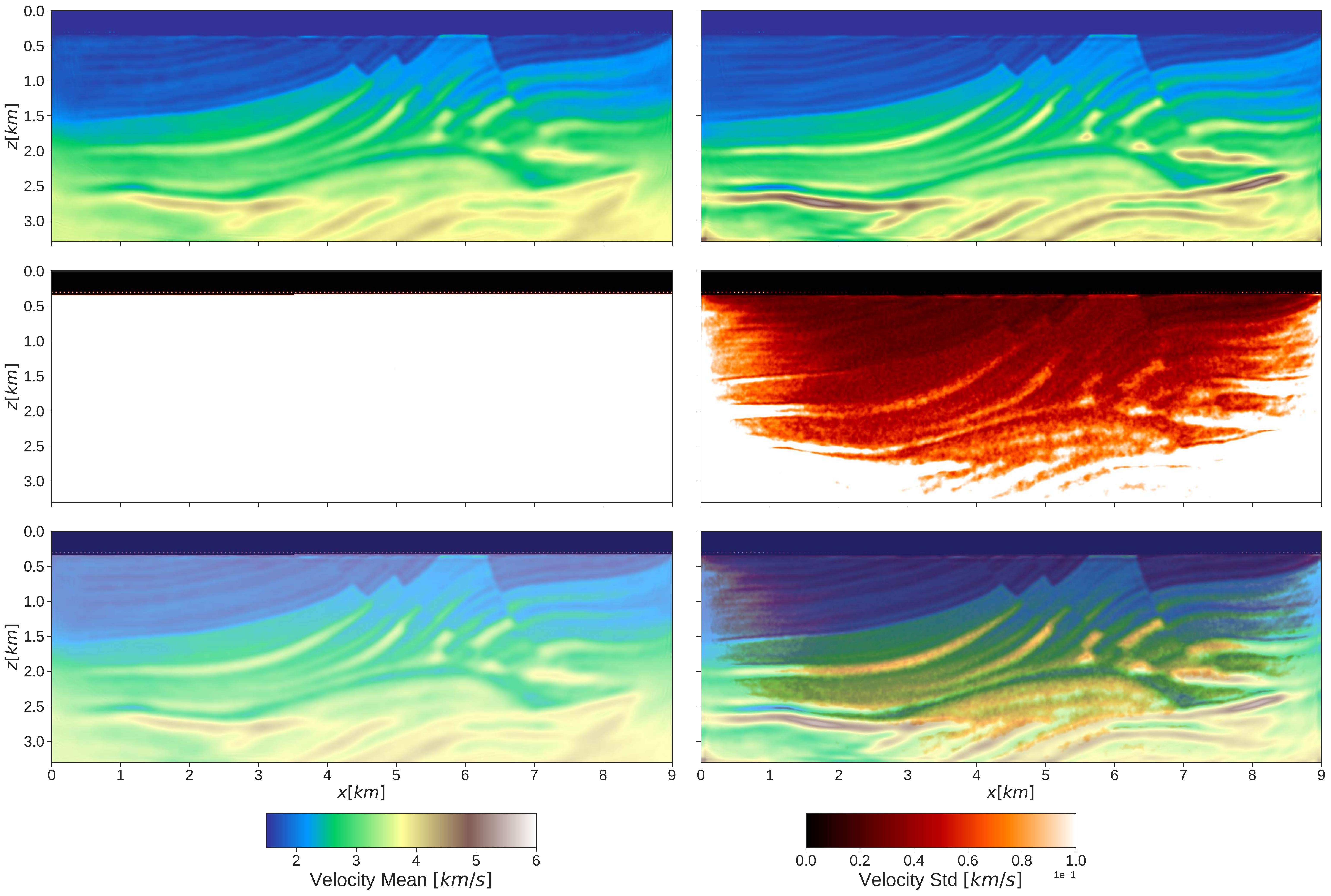}
  \caption{Marmousi model statistics with 100 particles. \textbf{Top row:} Initial and updated particles means. \textbf{Middle row:} Initial and updated particles standard deviations. \textbf{Bottom row:} Overlay of initial and updated particles means on their standard deviations, respectively.}
  \label{fig:rf-stats-100-particle}
\end{figure*}

The simulation with 50 particles took approximately 9 hours, while the 100 particles took about 20 hours. The statistics obtained from both simulations are illustrated in Figure~\ref{fig:rf-stats-50-particle} and~\ref{fig:rf-stats-100-particle}, respectively. Our numerical results suggest that by utilizing a small number of particles, 50 in this example, the standard deviation map, the second row of Figure~\ref{fig:rf-stats-50-particle}, presents qualitatively reliable information that honours the physics of wave propagation, similar to the one with a higher number of particles. However, the estimated magnitudes of uncertainty per grid point are biased due to the limited number of samples illustrated by the noisy pattern in the same image. Although the simulation with 100 particles comes with better accuracy in estimating the uncertainty of FWI, an accurate uncertainty map is commonly obtained by providing many SVGD particles (e.g., on the orders of $\sim10^{3}$ particles and above), but this will exhaust the computational resources and limit the applicability of uncertainty analysis in industrial-scale applications, especially for 3D seismic data.\\

\begin{figure*}[!htb]
  \centering
  \includegraphics[width=\textwidth]{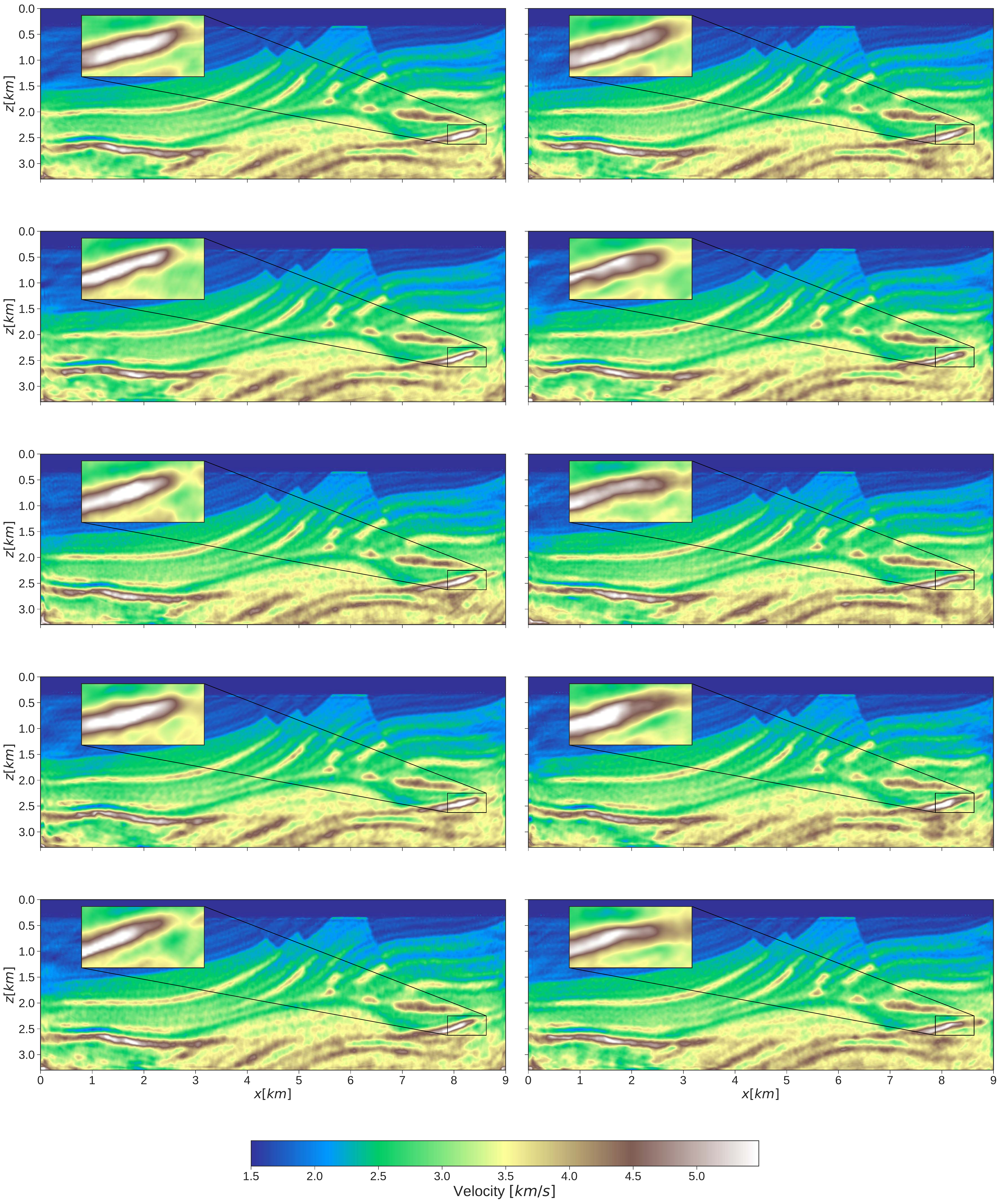}
  \caption{First ten realizations of the updated SVGD particles with GRF perturbation from simulation with 50 particles.}
  \label{fig:rf-post-50-particle}
\end{figure*}

Furthermore, the uncertainty map from both simulations indicates similar trends, where the regions with high model uncertainty are observed along the boundary and the bottom region of the Marmousi model. High standard deviation within those regions is expected as the data coverage does not allow for proper illumination of the deeper and on both sides of the model due to limitations of the geometry of acquisition and the physics of wave propagation within the subsurface. Such a phenomenon is illustrated vividly in Figure~\ref{fig:rf-post-50-particle}, where the FWI subsurface reconstruction within those high standard deviation regions shows high variability in the reflector positions due to disturbance in the low wavenumber component, signalling low information contributed by the observed data in correcting this phenomenon. We also obtain high uncertainty for high-velocity layers, reflecting the inherent low-resolution nature of waves propagating through them (long wavelengths). Overall, the resulting uncertainty map represents the holistic uncertainty derived from the observed data by incorporating the scattering and transmission components of FWI.\\

\begin{figure*}[!htb]
  \centering
  \includegraphics[width=\textwidth]{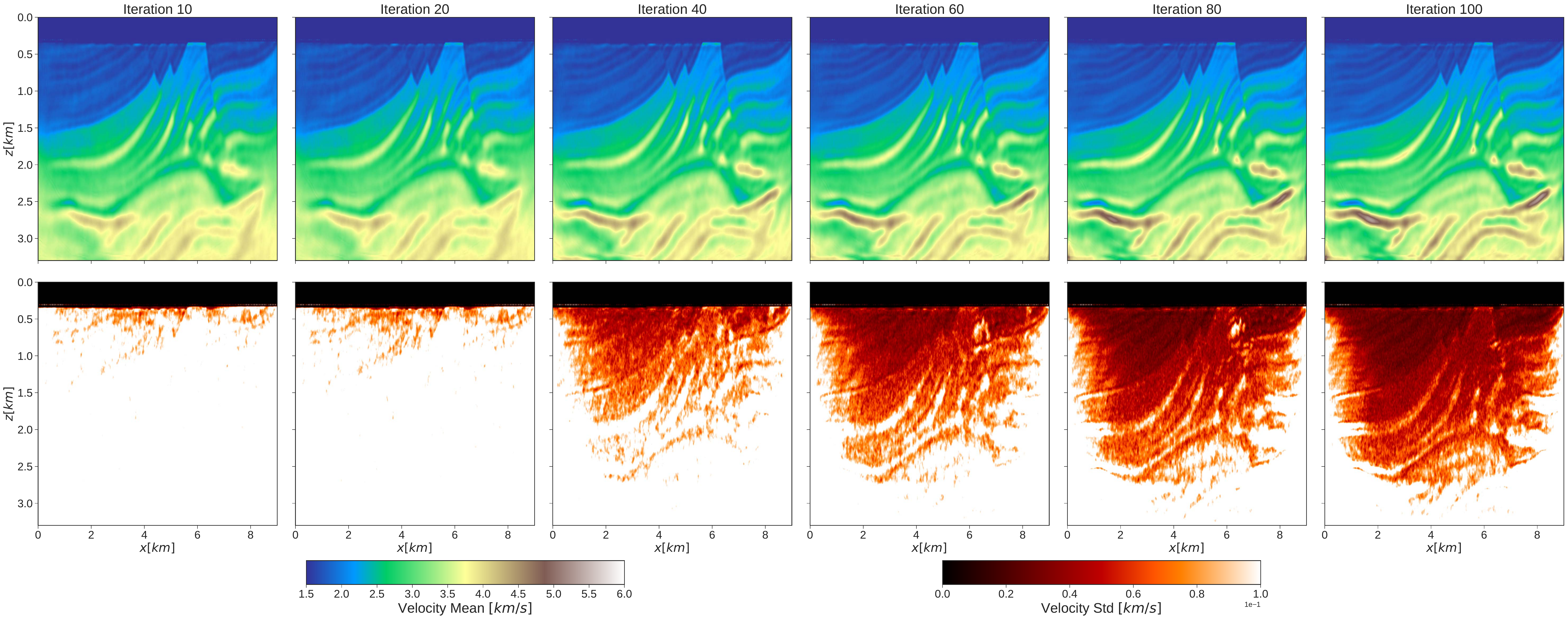}
  \caption{Marmousi model SVGD particles mean and standard deviation evolution over iterations with 50 particles.}
  \label{fig:rf-evo-50-particle}
\end{figure*}

\begin{figure*}[!htb]
  \centering
  \includegraphics[width=\textwidth]{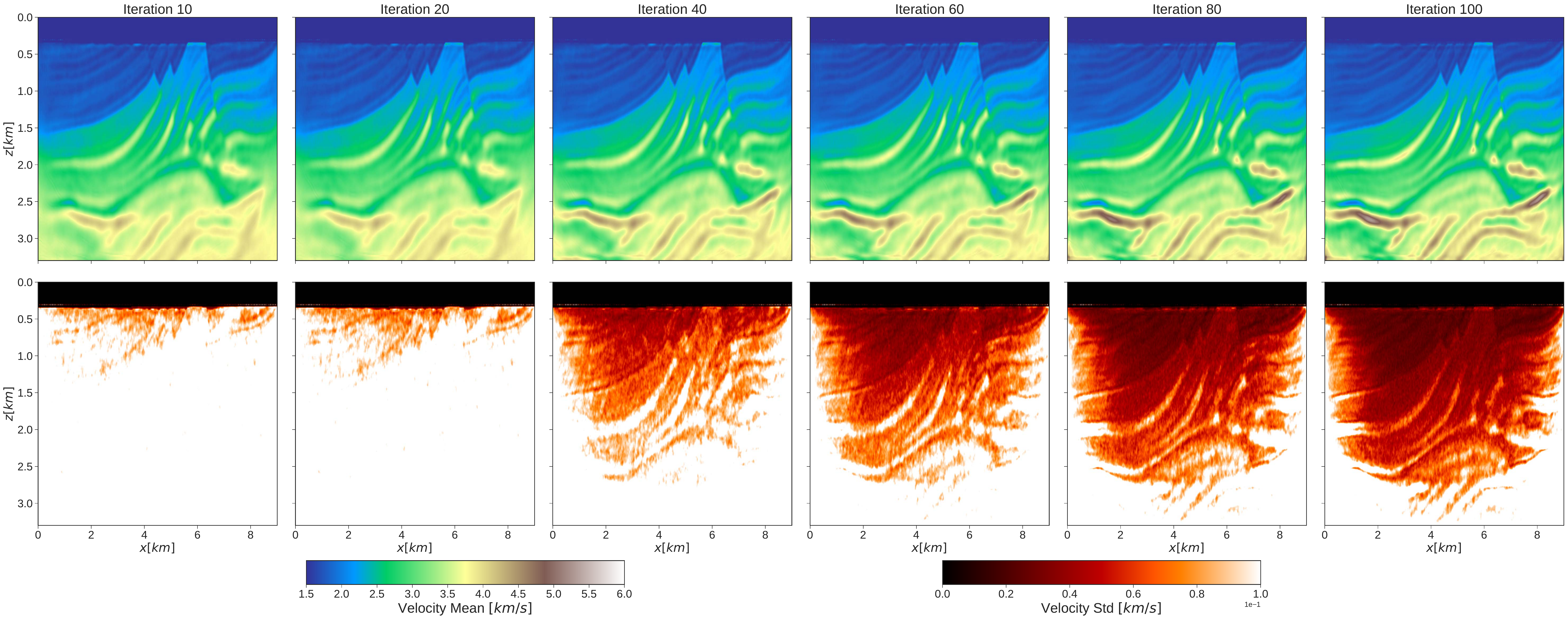}
  \caption{Marmousi model SVGD particles mean and standard deviation evolution over iterations with 100 particles.}
  \label{fig:rf-evo-100-particle}
\end{figure*}

Also, since we aim for computational saving, we explore the evolution of the SVGD particles' mean and standard deviation over iterations by illustrating them in Figure~\ref{fig:rf-evo-50-particle} and~\ref{fig:rf-evo-100-particle} for the 50 and 100 particles simulations, respectively. As observed, at early iterations, such as at iteration 40 (i.e., third columns in both figures), already provided qualitatively relevant information about the uncertainty of the inverted subsurface model at the cost of a noisy image. Nevertheless, the uncertainty map becomes more refined as iteration increases. This directly exemplified a promising path towards practically assisting decision-making at a reasonable computational cost for industrial-scale FWI applications.

\section{Discussion}
To perform uncertainty estimation at an industrial-scale FWI, the approach should be scalable and computationally efficient. However, an algorithm itself, such as SVGD, is insufficient to tackle industrial-scale applications. Here, the proposed frugal approach demonstrated credibility in estimating uncertainty in FWI by providing qualitatively relevant information about the inverted subsurface model from a small set of samples. It also demonstrated a promising path in practically assisting decision-making at a reasonable computational cost. In this section, we discuss the limitations of our approach from three perspectives--computational, statistical, and hyperparameter tuning. We also provide a discussion on addressing the uncertainty related to the transmission (i.e., low wavenumber) component, which controls wave propagation in FWI.

\subsection*{Computational and statistical aspects}
SVGD generally requires one forward and one adjoint operator evaluation per particle at each iteration; as such, the computational cost grows linearly with the number of particles. Furthermore, in a full Bayesian inference framework, to obtain a good accuracy in posterior sampling, SVGD requires at least more particles than the number of model parameters, which would limit its scalability on large-scale nonlinear inverse problems such as full waveform inversion (FWI)~\citep{zhang_variational_2020, zhang_3d_2023}. In this work, we alleviate this challenge by compromising the full Bayesian inference for scalability at industrial-scale FWI by deriving relevant statistical information from a small set of SVGD particles. This approach statistically provided biased statistics due to the limited number of samples. Despite this, we observed from the numerical examples that information about uncertainty in FWI remains relatively unchanged regardless of the limited number of particles used in SVGD. Nevertheless, the information derived from the proposed approach could only be used to infer the credibility of subsurface regions inverted by FWI, not for posterior distribution analysis or as an accuracy/error metric. 

\subsection*{SVGD hyperparameter tuning}
In SVGD, especially in the numerical examples shown, several hyperparameters influence its performance and efficiency, such as the choice of kernel and step size. Gorham and Mackey~\citep{gorham_measuring_2017} suggested using the inverse multiquadric (IMQ) kernel instead of the RBF kernel since it may not be efficient in high dimensions. However, this suggestion assumes a fixed bandwidth, which does not hold for the RBF kernel equipped with the median trick, which can better adapt to the samples at each iteration.  As for the step size, it controls the convergence of SVGD, where a proper step size decaying strategy would help the algorithms converge faster. In the demonstrated examples, we use cosine annealing~\cite{loshchilov_sgdr_2017} as the step size decaying strategy. A poor choice of step size would increase the computational cost as the number of FWI iterations would also increase, thus increasing the number of forward and adjoint operator evaluations. Note that, due to the nature of strong nonlinearity in FWI, the choice of step size would also influence the trajectory of the SVGD. We leave the study of the proper choice of step size to future works.

\subsection*{Uncertainty estimation for transmission component}
\begin{figure*}[!htb]
  \centering
  \includegraphics[width=\textwidth]{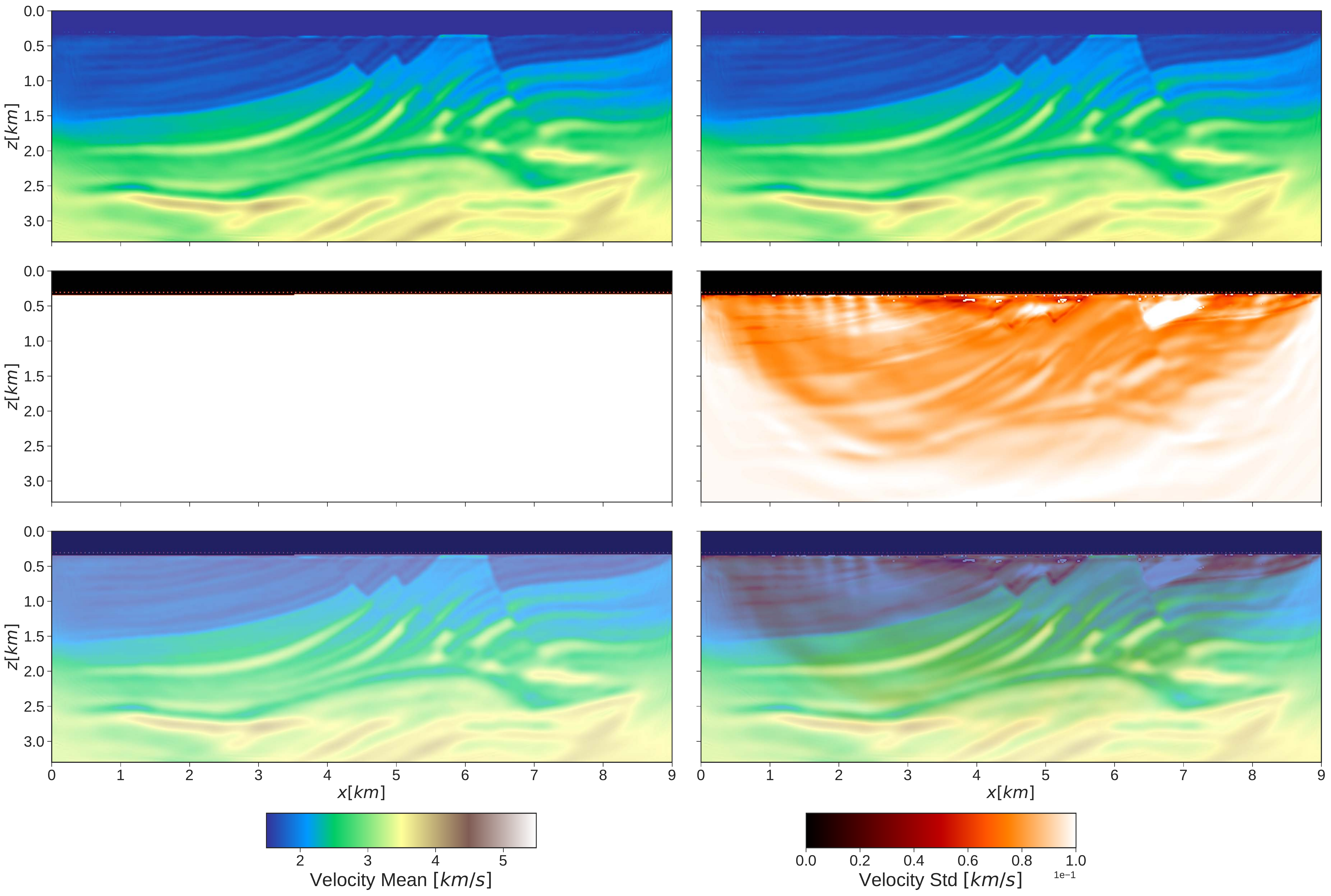}
  \caption{Marmousi model statistics with 100 particles from constant velocity perturbation model. \textbf{Top row:} Initial and updated particles means. \textbf{Middle row:} Initial and updated particles standard deviations. \textbf{Bottom row:} Overlay of initial and updated particles means on their standard deviations, respectively.}
  \label{fig:rf-stats-unif}
\end{figure*}

To illustrate the uncertainty which captures the transmission component in FWI, we performed an example with a similar setting as in the numerical examples section, but with constant velocity perturbations. Constant velocity perturbations admit an inversely proportional change in the traveltime courtesy of the eikonal equation. We sample 100 constant values from a uniform distribution, which results in constant velocity model perturbations between $\pm 50 m/s$. We later added these perturbations to the optimized velocity model obtained from the deterministic FWI procedure and considered them as the initial SVGD particles. Figure~\ref{fig:rf-stats-unif} illustrates the relevant statistics derived from the simulation. The updated uncertainty map in the second column of the middle row shows a similar trend as the recorded uncertainty from the GRF perturbation model, although it significantly varies in magnitude. The key takeaway from this result is that the constant velocity perturbation model highlights the kinematic component of the FWI (i.e., the tomography component) uncertainty. As a result, the areas of lower standard deviation (uncertainty) highlight the transmission path illumination even those reflected from interfaces, and thus, we observe a generally smaller region of low uncertainty compared to Figure~\ref{fig:rf-stats-100-particle}. This is a direct result of the fact that scattering can be recorded from more locations in the model than transmission, including closer to the edges of the model. Thus, using SVGD, we can overall isolate the uncertainties corresponding to the various components of FWI by altering the perturbations in the initial SVGD particles in correspondence with the model scale (i.e., wavenumber) they represent.


%
\section{Conclusions}
We proposed a frugal approach to estimate uncertainty in FWI through the Stein Variational Gradient Descent (SVGD) algorithm by utilizing a relatively small number of velocity model particles. By doing so, we can provide uncertainty maps, i.e., relative standard deviation maps, at a much lower cost, allowing for the potential for industrial-scale applications. We also introduced random field-based perturbations to the optimized velocity model from the deterministic FWI procedure as the SVGD initial particles. Such perturbations cover the scattering (i.e., high wavenumber) and the kinematic (i.e., low wavenumber) components of FWI; thus, the resulting uncertainty map represents the uncertainty of the FWI holistically. Our numerical results on the Marmousi model demonstrated credibility in estimating uncertainty in FWI by providing qualitatively relevant information about the inverted subsurface model from a relatively small set of samples, which directly exemplified a promising path in practically assisting decision-making at a reasonable computational cost for industrial-scale FWI applications.
\section*{Acknowledgment}
This publication is based on work supported by the King Abdullah University of Science and Technology (KAUST). The authors thank the DeepWave Consortium sponsors for their support.
\section*{Code Availability}
The data and accompanying codes that support the findings of this study are openly available at \href{https://github.com/DeepWave-KAUST/FWISVGD}{https://github.com/DeepWave-KAUST/FWISVGD}.


\printbibliography


\end{document}